\documentclass[sigconf]{acmart}

\usepackage[utf8]{inputenc}

\PassOptionsToPackage{hyphens}{url}\usepackage{hyperref}

\usepackage{csquotes}

\usepackage{graphicx}
\graphicspath{{images/}}

\usepackage{multirow}

\usepackage{makecell}

\usepackage{array}

\usepackage{dblfloatfix}

\usepackage{enumitem}

\copyrightyear{2024} 
\acmYear{2024} 
\setcopyright{rightsretained} 
\acmConference[ICSE-SEIS'24]{Software Engineering in Society}{April 14--20, 2024}{Lisbon, Portugal}
\acmBooktitle{Software Engineering in Society (ICSE-SEIS'24), April 14--20, 2024, Lisbon, Portugal}
\acmDOI{10.1145/3639475.3640111}
\acmISBN{979-8-4007-0499-4/24/04}

\begin{document}

\title{A Synthesis of Green Architectural Tactics for ML-Enabled Systems}

\author{Heli Järvenpää}
\orcid{0009-0000-8669-3079}
\affiliation{
    \institution{Vrije Universiteit Amsterdam}
    \city{Amsterdam}
    \country{The Netherlands}
}
\email{h.m.jarvenpaae@student.vu.nl}

\author{Patricia Lago}
\orcid{0000-0002-2234-0845}
\affiliation{
    \institution{Vrije Universiteit Amsterdam}
    \city{Amsterdam}
    \country{The Netherlands}
}
\email{p.lago@vu.nl}

\author{Justus Bogner}
\orcid{0000-0001-5788-0991}
\affiliation{
    \institution{Vrije Universiteit Amsterdam}
    \city{Amsterdam}
    \country{The Netherlands}
}
\email{j.bogner@vu.nl}

\author{Grace Lewis}
\orcid{0000-0001-9128-9863}
\affiliation{
    \institution{Carnegie Mellon Software Engineering Institute}
    \city{Pittsburgh, PA}
    \country{USA}
}
\email{glewis@sei.cmu.edu}

\author{Henry Muccini}
\orcid{0000-0001-6365-6515}
\affiliation{
    \institution{FrAmeLab, University of L'Aquila}
    \city{L'Aquila}
    \country{Italy}
}
\email{henry.muccini@univaq.it}

\author{Ipek Ozkaya}
\orcid{0000-0002-7336-4775}
\affiliation{
    \institution{Carnegie Mellon Software Engineering Institute}
    \city{Pittsburgh, PA}
    \country{USA}
}
\email{ozkaya@sei.cmu.edu}

\renewcommand{\shortauthors}{Järvenpää et al.}

\begin{abstract}
The rapid adoption of artificial intelligence (AI) and machine learning (ML) has generated growing interest in understanding their environmental impact and the challenges associated with designing environmentally friendly ML-enabled systems.
While Green AI research, i.e., research that tries to minimize the energy footprint of AI, is receiving increasing attention, very few concrete guidelines are available on how ML-enabled systems can be designed to be more environmentally sustainable.
In this paper, we provide a catalog of 30 green architectural tactics for ML-enabled systems to fill this gap.
An architectural tactic is a high-level design technique to improve software quality, in our case environmental sustainability.
We derived the tactics from the analysis of 51 peer-reviewed publications that primarily explore Green AI, and validated them using a focus group approach with three experts.
The 30 tactics we identified are aimed to serve as an initial reference guide for further exploration into Green AI from a software engineering perspective, and assist in designing sustainable ML-enabled systems.
To enhance transparency and facilitate their widespread use and extension, we make the tactics available online in easily consumable formats.
Wide-spread adoption of these tactics has the potential to substantially reduce the societal impact of ML-enabled systems regarding their energy and carbon footprint.
\end{abstract}



\maketitle

\textbf{Lay Abstract:}
Machine learning (ML) is a technology field that wants to provide software with functionality similar to human-like intelligence, e.g., for understanding text or describing images.
However, creating and using systems with ML needs a lot more computing power than non-ML systems, which is bad for the environment.
Companies therefore need concrete advice on how they can create ML systems that are environmentally sustainable.
In this paper, we provide a catalog of 30 green architectural tactics for these systems.
An architectural tactic is a high-level design technique to improve software quality, in our case environmental sustainability.
To achieve this, we analyzed 51 scientific papers and later discussed with 3 experts to improve and extend our catalog.
If many companies start using these tactics, the energy footprint of systems with ML can be greatly reduced.

\section{Introduction}
\label{s:intro}
Artificial intelligence (AI) and machine learning (ML) have shown significant potential in digital innovations, with a growing number of different ML applications expanding across a wide spectrum of industries, from healthcare to agriculture and management~\cite{pallathadka2023impact}.
This rapid growth of ML applications has also drawn attention to its environmental footprint.
Several studies have evaluated the carbon emissions of ML~\cite{anthony2020carbontracker, lacoste2019quantifying}.
It is widely acknowledged that training and using ML models at scale is computationally demanding, which leads to greenhouse gas emissions.
For example, training a typical transformer-based natural language processing (NLP) model produces greenhouse gas emissions equivalent to five average cars in their lifetime~\cite{strubell2019energy}.
These considerations have led to new concepts such as \textit{Green AI} and \textit{Sustainability of AI}.
Traditional AI has aimed to achieve high accuracy while disregarding energy efficiency, but Green AI emphasizes the environmental footprint of AI and focuses on minimizing computation while still producing accurate results~\cite{schwartz2019green}.
\textit{Sustainability of AI} refers to the environmental impact of the AI model itself, and highlights the responsible development and use of AI-based systems in a way that minimizes negative impact on the environment and society~\cite{van2021sustainable}.

ML models present unique challenges for the development of ML-enabled systems due to their data-dependent nature, dynamic behavior, and lack of transparency~\cite{lewis2021software}.
These unique characteristics of ML have affected the fields of software engineering and software architecture, leading to an emerging focus on \textit{software architecture for ML-enabled systems} or SA4ML.
SA4ML aims to develop principles, practices, and tools for improving the development of ML-enabled systems~\cite{muccini2021software}.
Papers on patterns and tactics for ML-enabled systems have started to appear~\cite{washizaki2022software,heiland_design_2023}, but they concentrate primarily on developing effective ML-enabled systems while lacking the environmental perspective.

Despite the increasing interest in the environmental impact of ML~\cite{verdecchia_systematic_2023}, it is still challenging to obtain a comprehensive overview of the recommended best practices for developing green ML-enabled systems.
To gain a deeper understanding of the effectiveness and usefulness of different practices, we explore the intersection of Green AI and software architecture by identifying \textit{green architectural tactics for ML-enabled systems}.
We achieve this through a literature-based synthesis and a subsequent focus group review with software architecture experts working on SA4ML to validate and extend the initial collection of tactics.
In our study, we address the following research questions:

\begin{itemize}
  \item \textbf{RQ1}: Which green architectural tactics for ML-enabled systems can we synthesize from scientific literature?
  \item \textbf{RQ2}: How do SA4ML experts perceive our synthesized collection of tactics?
\end{itemize}

Our contribution is a collection of 30 green architectural tactics for ML-enabled systems.
To structure the collection, we organize the tactics into six categories that map to different aspects of ML-enabled systems development.
Given that a tactic is a design technique used to improve a specific quality attribute ~\cite{bass2021software}, we expect our contribution to guide practitioners to make more environmentally sustainable decisions when engineering ML-enabled systems.

\begin{figure*}[ht!]
    \centering
    \includegraphics[width=0.85\textwidth]{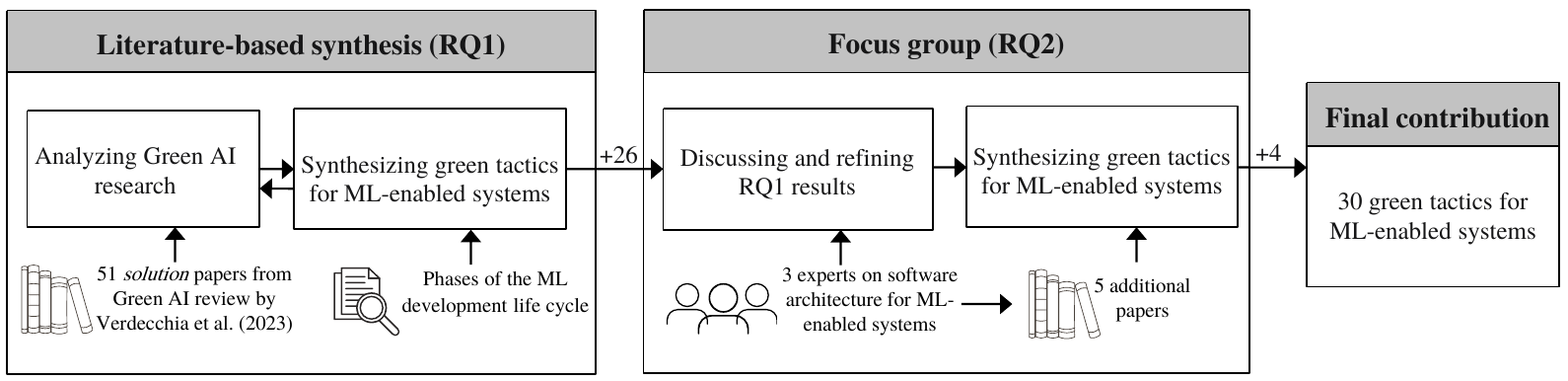}
    \vspace{-5pt}
    \caption{Study Design and Execution}
    \label{study-design}
\end{figure*}

\section{Background and Related Work}
\label{s:background}
We first provide an overview of important concepts such as ML-enabled systems, environmental sustainability, and architectural tactics, and then discuss related work in the area.

\subsection{ML-Enabled Systems}
ML is a sub-field of AI that concentrates on developing algorithms capable of learning from data~\cite{sarker2021machine}.
ML models are trained on input data to produce a desired outcome without being explicitly programmed to do so: they automatically identify connections in the data and learn complex patterns~\cite{ElNaqa2015}.
These learned patterns can then be used to make predictions or decisions for new, previously unseen data.
According to \citet{amershi2019software}, building an ML model usually follows a specific development process with feedback loops: understanding business goals, preparing the data, designing the model, training the model, testing the model, deploying the model in the production environment, and finally managing and monitoring it during usage.

While ML models can be used for stand-alone data science or analytics activities, they are also frequently embedded into larger software systems.
The term \textit{ML-enabled system} therefore refers to a software system that includes at least one ML component~\cite{lewis2021software}.
Due to the special characteristics of machine learning, integrating ML components into software systems comes with additional challenges in terms of software architecture and design, e.g., strong dependencies on training data, lack of transparency for ML component behavior, and abrupt changes in the prediction quality of ML components~\cite{lewis2021software}.
Ensuring quality is therefore an important and complicated activity for ML-enabled systems, with many quality attributes (QAs) to consider~\cite{gezici2022systematic}.
The unique characteristics of ML model and system development have also given rise to special system-level QAs, which include, e.g., prediction accuracy, explainability, monitorability, sustainability, prediction cost, training cost, and training latency~\cite{horkoff2019non, vogelsang2019requirements, siebert2020towards, pons2019priority}.

\subsection{Environmental Sustainability of ML}
Holistically evaluating the impact of an ML-enabled system on environmental sustainability is challenging because the system goes through several stages during its life cycle, from ML model development to deployment, maintenance, and evolution of the ML-enabled system.
However, the negative environmental impact of ML is usually directly related to the required high computational power, i.e., the amount of energy required to train and run ML models~\cite{kaack2022aligning}.
Computational power is typically evaluated via energy consumption measured in joules (J) or kilowatt-hours (kWh), computing power via floating-point operations per second (FLOPS), GPU/CPU utilization as a percentage, GPU/CPU hours, response time, or carbon emissions~\cite{Henderson2020}.
Additional factors beyond software-related computational power like hardware manufacturing, transportation, or e-waste are harder to quantify and outside the study scope.

The development stage of ML models is usually regarded as the most energy-intensive phase within the life cycle of ML-enabled systems~\cite{kaack2022aligning}, especially for systems using generative AI, such as in the form of large language models (LLMs)~\cite{jovanovic_generative_2022}, since these models require large amounts of data.
However, increasing the number of data points is also used as a general strategy to train more accurate ML models, which makes data preprocessing and model training consume more energy~\cite{wu2022sustainable}.
Training data may also be complex and multidimensional, which is why data cleaning, labeling, and feature engineering can be energy-intensive~\cite{wu2016big}.
Along with the increased amount of data, ML model size and complexity also increase.
As a result, the model training and tuning phases are very energy-intensive, especially if multiple rounds of testing are required~\cite{schwartz2019green}.
For example, many deep learning models are often trained for long periods of time, ranging from dozens to thousands of hours, sometimes using powerful but energy-intensive hardware such as GPUs or other processing units optimized for ML workloads~\cite{berriel2017monthly}.
While ML training receives the majority of attention, each development stage in the ML development process has its own concerns and opportunities related to energy consumption and environmental sustainability.
Therefore, evaluating the sustainability of ML requires a holistic view of the ML life cycle~\cite{wu2022sustainable}.

\subsection{Green Architectural Tactics}
Synthesizing generalizable design decisions to achieve certain QAs is an important practice in the software architecture community to enable the reuse of architecture knowledge.
One concept in this space is an \textit{architectural tactic} (or \textit{tactic} in short form), i.e., a high-level design decision focusing on achieving a single QA for a software system~\cite{bass2021software}.
Tactics differ from \textit{design patterns}~\cite{Gamma1994}.
Patterns typically are more concrete and cover multiple decisions, tactics, and QAs, often through trade-offs that impact some QAs positively and others negatively.
In this sense, tactics are high-level building blocks from which patterns are composed, but choosing an architecture pattern can also guide the selection of complementary architectural tactics~\cite{harrison_how_2010}.
\citet{harrison_how_2010} discuss the relationship between tactics and patterns in greater detail.

Architectural tactics often improve one primary QA and are typically organized in QA-related collections such as security or performance tactics~\cite{marquez_architectural_2023}.
However, tactics may also have a positive influence on other secondary QAs, even though these QAs may not be the main focus of the tactic.
For example, a tactic aimed at improving performance via less resource-intensive computation and faster response times will often also improve energy efficiency.
In the context of ML-enabled systems, we focus on \textit{green} tactics, i.e., tactics that have a positive impact on environmental sustainability, usually in the form of improving energy or carbon efficiency.
All architectural tactics are design decisions that positively impact a QA of a system or part of a system, and can  manifest themselves within different scopes.
For example, we later introduce the tactics \textit{Remove redundant data} (T2) and \textit{Decrease model complexity} (T9).
These tactics do not focus on the complete software system, but rather focus solely on the data (T2) and model (T9) as their scope.
This interpretation is in line with the definition of software architecture as the set of significant design decisions~\cite{Jansen2005}.

\subsection{Related Work}
Architectural tactics are a popular medium to convey design knowledge in the software engineering research community.
A recent mapping study by \citet{marquez_architectural_2023} identified a total of 91 primary studies.
However, the authors criticized that most studies use the concept of tactics in a way that is not in line with the original definition, and that it is often unclear from which data sources the tactics were synthesized.
They also state that they found little evidence of the use of architectural tactics in industry.

While sustainability or energy efficiency are not explicitly covered by the primary studies discussed by \citet{marquez_architectural_2023}, some authors have proposed green tactics for different domains.
\citet{procaccianti_catalogue_2014} and \citet{vos_architectural_2022} synthesized tactics for optimizing the energy efficiency of software running in the (public) cloud.
Similarly, \citet{chinnappan_architectural_2021} and \citet{malavolta_mining_2021} conceptualized tactics for energy-aware robotics software.
While these could be situationally useful for ML-enabled systems in such domains, it is important to have tactics that take the specifics of ML into account.

By now, many publications exist on the general topic of Green AI.
A recent review by \citet{verdecchia_systematic_2023} collected 98 studies in this fast-growing field.
However, the majority of these publications concentrate on understanding the mechanisms related to the energy efficiency of ML and do not provide their findings in an easily accessible and actionable form.
Moreover, the existing techniques are distributed among many different publications, making their holistic consideration difficult.
In this sense, a collection of concrete tactics that practitioners can use to make ML-enabled systems more sustainable does not currently exist.
One work that starts to produce actionable guidance is a study by \citet{shanbhag2022towards}: they used the concept of design patterns to synthesize eight \enquote{energy patterns} for deep learning projects from an analysis of Stack Overflow posts.
The patterns were subsequently validated with a questionnaire survey with 14 practitioners.
While their catalog is a promising step in the right direction, our study aimed for a broader scope and coverage: we included all forms of ML, derived our tactics from scientific literature, and took a holistic look at the life cycle of ML-enabled systems.
This ultimately led to the 30 green tactics we synthesized, which also include the 8 energy patterns from \citet{shanbhag2022towards}.

\begin{figure*}[b!]
    \centering
    \includegraphics[width=0.85\textwidth]{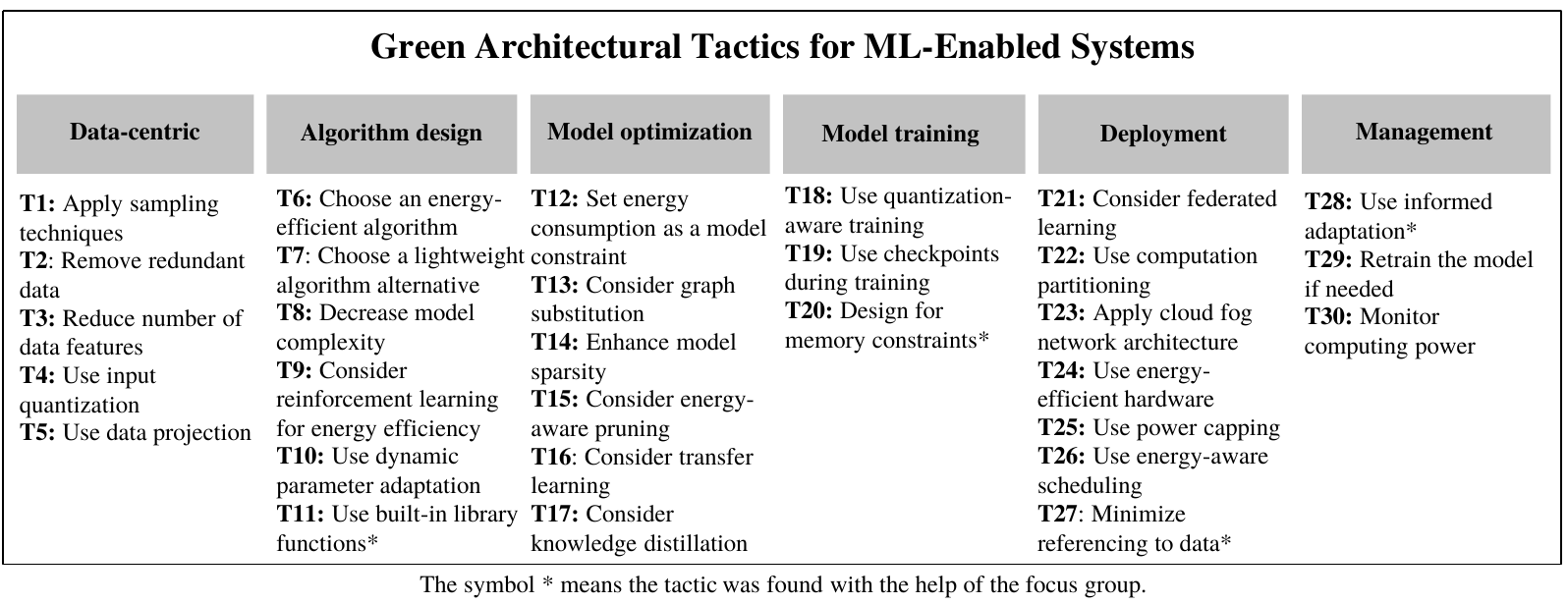}
    \caption{Catalog of the 30 Synthesized Green Architectural Tactics for ML-Enabled Systems}
    \label{tactics:all}
\end{figure*}

\section{Study Design and Execution}
\label{s:study-design}
We designed our study with a qualitative research approach organized in two steps (see Fig.~\ref{study-design}).
In the first step, we used scientific literature on Green AI to synthesize green architectural tactics for ML-enabled systems (RQ1).
To structure the tactics, we formed categories and iteratively refined the collection.
Once the collection reached sufficient maturity, we carried out the second step, in which we conducted a focus group with three SA4ML experts (RQ2).
This allowed us to validate and further refine the collection of tactics.

A focus group is a qualitative method to collect opinion-based data about a specific topic through a systematic discussion with participants knowledgeable about the topic~\cite{shull_focus_2008}.
The number of participants needs to be small enough for everyone to contribute, and a researcher has to carefully moderate the exchange, typically by following a predefined protocol with questions.
Based on such a group setting, exchanges between participants can stimulate new ideas and lead to stronger consensus.
Using the focus group results, the collection was substantially improved and extended. 

\subsection{Literature-Based Synthesis (RQ1)}
\label{s:synthesis}
The data collection for the first part of the study was based on a recent systematic literature review on Green AI by \citet{verdecchia_systematic_2023}.
They queried the publisher-agnostic search engines Google Scholar, Scopus, and Web of Science, and complemented this with an iterative snowballing process, leading to a total of 98 primary studies.
These studies were categorized into \textit{observational}, \textit{solution}, and \textit{position} articles.
The \textit{solution} category included papers providing techniques or tools to address Green AI issues, e.g., to improve the energy efficiency of ML models or components.
Because these papers represent promising sources for architectural tactics, we used all 51 papers from this category as the start for our synthesis.
Papers from the other two categories (\textit{observational} and \textit{position}) did not contain relevant information to synthesize actionable techniques.

Following the guidelines of \citet{bowen2009document}, all 51 selected papers were evaluated with qualitative data analysis methods, mainly \textit{document analysis} and \textit{content analysis}.
As a first step, the main researcher performed a document analysis for each paper that included skimming, reading, and interpretation, which was followed by a content analysis to identify and synthesize green tactics for ML-enabled systems.
This content analysis was guided by the \enquote{Awesome Tactic} template~\cite{awesome-template} from the Archive of Awesome and Dark Tactics (AADT).
AADT is a curated, open-source knowledge base that provides information on tactics for various domains that can have either a positive or negative impact on software quality.
The AADT template for tactics is grounded in foundational literature about architectural tactics and includes fields such as \textit{tactic intent}, \textit{participant}, \textit{context}, \textit{primary QA}, \textit{secondary QAs}, and \textit{measured impact}.
Using descriptive coding methods~\cite{miles2014qualitative}, the synthesized tactics were organized into categories representing different aspects of the development life cycle of ML-enabled systems.
The analysis took place in an iterative manner, with frequent reviews and exchanges between researchers in the spirit of continuous refinement.

\subsection{Focus Group (RQ2)}
In the second part of the study, we conducted a focus group with three experts in software architecture for ML-enabled systems (SA4ML).
As is common with focus groups~\cite{shull_focus_2008}, we used purposive sampling to recruit participants from our network whom we knew to be very knowledgeable and experienced about the topic.
All three experts were in senior, research-related roles with extensive expertise in software architecture, ML-enabled systems, and software sustainability.
One participant worked at a university and two participants at a research institute.
The experts did not have ML engineering roles in industry, but all of them had extensive knowledge of the state of ML engineering in industry through frequent collaborations and projects with companies or public organizations that developed or acquired ML-enabled systems.
Two authors were present during the focus group as moderators.
The session took place as a video call of roughly 75 minutes, which was recorded for later analysis with everyone's permission.

Following a protocol with questions, the moderators presented the collection of synthesized green tactics and encouraged an open discussion to receive feedback.
One aim was to validate the tactics' soundness, relevance, and applicability to industry scenarios.
Additionally, the protocol included open questions to find out if important tactics applied in industry were missing.
This led to an in-depth discussion about the tactics and their categorization.
The participants suggested several additional publications to extend the collection.
These papers were included in the research to extract possible tactics from them.

Afterward, the recording of the focus group was analyzed to identify the key themes of the discussion.
The main topics were written down and used to improve the tactics identified in the first step of the study.
All the feedback received about the identified tactics was considered, and, if applicable, the tactics were modified accordingly.
The additional papers suggested by the participants were analyzed using the same document and content analysis approach described in Section \ref{s:synthesis} to synthesize additional green tactics for ML-enabled systems.
In particular, the focus group added the following contributions:

\textbf{Extension of the catalog:} as mentioned, the experts suggested additional papers that led to the identification of four new tactics.

\textbf{Refinements of descriptions and categories:} the descriptions and categorization of the tactics benefited from the insights provided by the experts, which observed that: \textit{(i)} while all tactics are architecture-relevant, many are more model-focused, i.e., related to the ML model rather than the system as a whole.
Overall, the tactics are quite diverse and focus on data management, process aspects, or design vs. implementation aspects to various degrees; \textit{(ii)} most tactics target energy efficiency as a primary QA, but some also focus on other QAs or include important trade-offs with secondary QAs; \textit{(iii)} some tactics are specific to certain ML techniques and, as such, are not universally applicable to achieve energy efficiency.
This led to renaming some tactics from \enquote{Use \dots} to \enquote{Consider \dots}. For example, \textit{Consider transfer learning} (T16) can reduce energy consumption, but it is not always applicable; so one might say that, if applicable, T16 is preferred to increase energy efficiency.

\textbf{Observations and reflections for future work:} the experts provided interesting observations and reflections that point to possible follow-up work. These are discussed in Section \ref{s:discussion}.

\subsection{Threats to Validity}
According to \citet{Verdecchia2023}, threats to validity are often considered as an afterthought in the SE community, and trade-offs among threats to validity are often neglected.
For this study, we prioritized the soundness and relevance of the collection.
Therefore, we consciously accepted limitations in our study design that decreased external validity, e.g., completeness and generalizability, but improved internal validity, e.g., soundness and consistency.

In this regard, one threat to validity is that our data collection relied on the literature review of \citet{verdecchia_systematic_2023}.
While the authors followed a sound protocol with extensive snowballing, their initial search terms were focused on Green AI in general, and did not include terms related to ML-enabled systems or software architecture.
This could have led to some papers relevant to green architectural tactics for ML-enabled systems not being found.
However, the subsequent focus group partly mitigated this threat by suggesting additional literature.

The focus group method can also be prone to some threats to validity.
Because it involves relatively few participants, the generalizability of the results can be impacted.
However, this also allows an in-depth discussion of the topic with rich, qualitative reflections.
Additionally, we ensured that the participants were prestigious experts on the topic, with extensive experience and expertise.

Lastly, qualitative research can be prone to subjective biases that can influence the results.
To mitigate this, the collection of tactics was not only validated by the focus group, but also extensively reviewed and refined by the research team until consensus was reached.
During these iterative refinements, great care was taken to preserve clarity and consistency, e.g., through proofreading of a refined tactic by someone else.
Three researchers were involved in the in-depth synthesis and refinement of the tactics, with the rest providing high-level feedback.
All in all, we do not claim completeness for our final collection of tactics, but we demonstrate a representative list of actionable and relevant tactics that can support practitioners in creating more environmentally sustainable ML-enabled systems.

\section{Results}\label{s:results}
\begin{figure}[H]
    \centering
    \includegraphics[width=0.85\columnwidth]{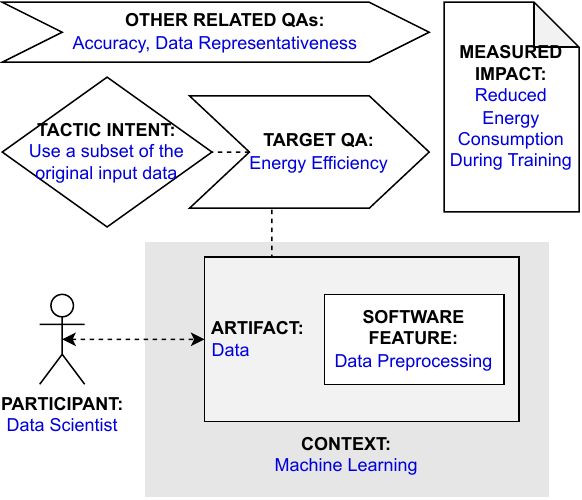}
    \caption{Tactic Template for \textit{Apply Sampling Techniques} (T1)}
    \label{fig:tactic-template-example}
    \vspace{-7pt}
\end{figure}

The initial literature-based synthesis before the focus group discussion resulted in 26 tactics.
Based on the focus group, we validated these tactics and identified four additional tactics via the literature suggested by the experts (T11, T20, T27, and T28).
Our final collection therefore consists of 30 green architectural tactics for ML-enabled systems.
They are organized into six categories: \textit{data-centric}, \textit{algorithm design}, \textit{model optimization}, \textit{model training}, \textit{deployment}, and \textit{management}.
These categories correspond to the different types of artifacts and phases relevant to the life cycle of ML-enabled systems so that tactics can be selected based on current status or for certain parts of the system.
A summary of the resulting tactics is shown in Fig.~\ref{tactics:all}.
The collection is also available in the online repository that accompanies this paper\footnote{\url{https://doi.org/10.5281/zenodo.8349208}} and in the Archive of Awesome and Dark Tactics.\footnote{\url{https://s2group.cs.vu.nl/AwesomeAndDarkTactics/catalog}}
To illustrate and explain the used tactic template, we present one example in full detail below (see Fig.~\ref{fig:tactic-template-example} for the visual template).
The remaining tactics will be presented in short summaries in their respective category subsections.

The tactic \textit{Apply sampling techniques} (T1) primarily intends to improve energy efficiency (\textit{Target QA}), but also impacts accuracy and data representativeness (\textit{Other Related QAs}).
In the template, the impact for the \textit{Target QA} is always positive, but for \textit{Other Related QAs}, it can also be negative, e.g., T1 can reduce accuracy.
The \textit{Tactic Intent} summarizes the general technique that is applied, i.e., \enquote{use a subset of the original input data} for T1.
In the \textit{Context} of \enquote{machine learning}, the targeted \textit{Artifact} of T1 is \enquote{data}.
Other tactics target, e.g., models, training algorithms, or deployment infrastructure instead.
The related \textit{Software Feature} of T1 is \enquote{data preprocessing}, with other tactics focusing on, e.g., model training or inference.
T1 is usually applied by the \textit{Participant} \enquote{data scientist}.
For tactics with empirical evidence, the \textit{Measured Impact} is reported, i.e., \enquote{reduced energy consumption during training} for T1.
The textual template also includes the references to the respective studies.

\subsection{Data-Centric}
This category includes tactics that process input data for ML models to promote energy efficiency, shown as tactics T1-T5 in Table~\ref{data-tactics}.

\begin{table}[ht]
\centering \small 
\caption{Data-Centric Green Tactics for ML-Enabled Systems}
\begin{tabular}{>{\raggedright\arraybackslash}p{1.9cm}>{\raggedright\arraybackslash}p{3cm}>{\raggedright\arraybackslash}p{1.5cm}>{\raggedright\arraybackslash}p{0.8cm}}
\hline
\textbf{Tactic} & \textbf{Description} & \textbf{Target QA} & \textbf{Source} \\
\hline T1: Apply sampling techniques & Use a smaller subset of the original input dataset & Energy efficiency & \cite{verdecchia2022data}\cite{Wang2019} \\
\hline
T2: Remove redundant data & Detect and remove redundant data from the original input data & Energy efficiency & \cite{Ang2022}\cite{dhabe2021data} \\
\hline
T3: Reduce number of data features & Reduce the number of input data features used & Energy efficiency & \cite{verdecchia2022data} \\
\hline
T4: Use input quantization & Convert input data to smaller precision & Accuracy* & \cite{abreu2022}\cite{Kim2021} \\ 
\hline
T5: Use data projection & Project data into a lower-dimensional embedding & Performance* & \cite{Rouhani2016Aug}\\ 
\hline
\multicolumn{4}{l}{\small{The * means energy efficiency was considered a secondary QA}} \\
\end{tabular}
\label{data-tactics}
\end{table}

\vspace{-3pt}

\textbf{T1: Apply sampling techniques}.
The size of the input data has a positive correlation with the energy consumption of computing~\cite{ang2022characterizing}.
Therefore, reducing the size of the input data can have a positive impact on the energy efficiency of ML.
Sampling techniques, such as simple random sampling or systematic sampling, offer different approaches to selecting a subset of the input data.
\citet{verdecchia2022data} employed stratified sampling to reduce the number of data points.
Stratified sampling means randomly selecting data points from homogeneous subgroups of the original dataset.
This technique resulted in savings in energy consumption~\cite{verdecchia2022data}.

\textbf{T2: Remove redundant data}.
This tactic aims to decrease the size of the input data, which consequently reduces the size of the model.
Identifying and removing redundant data for ML models can decrease computing time, the number of computations, energy consumption, and memory space.
Redundant data refers to those data points that do not contribute significantly to the accuracy of the model.
Therefore, removing these unimportant data points does not sacrifice much accuracy.
Removing redundant data reduces the energy consumption of training and inference~\cite{dhabe2021data, sun2020}.

\textbf{T3: Reduce number of data features.} A large number of data features in the ML model can lead to high computing power requirements during training and inference.
Typically, ML scenarios involve a huge number of features or variables that describe the input data.
However, not all of these features are necessary for the model to make accurate predictions.
Therefore, reducing the number of input data features can lead to improved energy efficiency while maintaining accuracy.
This reduction can be achieved by selecting only a subset of all the available data features~\cite{verdecchia2022data}.

\textbf{T4: Use input quantization.} Input quantization in ML refers to converting the data to a smaller precision, e.g., reducing the number of bits used to represent the data.
According to a study by \citet{abreu2022}, using 10-bit precision is sufficient for achieving accuracy in ML models, and using more does not contribute to accuracy.
Therefore, using higher precision in this case is a waste of resources.
Data quantization can also be used in federated learning to reduce energy consumption and memory access~\cite{Kim2021}.
Using precise data values through input quantization can even have a positive impact on the accuracy of the ML model by reducing overfitting.

\textbf{T5: Use data projection}: Data projection means transforming data into a lower-dimensional embedding.
Reducing the dimensionality of input data shrinks the dimensionality of deep neural networks (DNN), which leads to improved performance of the model.
Using data projection as a preprocessing step can result in energy improvements without sacrificing performance or accuracy~\cite{Rouhani2016Aug}.

\subsection{Algorithm Design}
Tactics in the algorithm design category refer to decisions that are made when designing the ML model.
This category consists of six different tactics (T6-T11) that are shown in Table \ref{tactics-algorithm}.

\begin{table}[ht]
\caption{Green Tactics Related to Algorithm Design}
\centering \small 
\begin{tabular}{>{\raggedright\arraybackslash}p{1.9cm}>{\raggedright\arraybackslash}p{3cm}>{\raggedright\arraybackslash}p{1.5cm}>{\raggedright\arraybackslash}p{0.8cm}}
\hline
\textbf{Tactic} & \textbf{Description} & \textbf{Target QA} & \textbf{Source} \\
\hline T6: Choose an energy-efficient algorithm & Choose the most energy-efficient algorithm that achieves sufficient level of accuracy & Energy efficiency & \cite{kaack2022aligning}\\
\hline
T7: Choose a lightweight algorithm alternative & If possible, choose lighter alternatives of existing algorithms & Energy efficiency & \cite{sorbaro2020}\\
\hline
T8: Decrease model complexity & Decrease the complexity of an ML model & Energy efficiency & \cite{abreu2020}\cite{morotti2021}\\
\hline
T9: Consider reinforcement learning for energy efficiency & Use reinforcement learning to optimize energy efficiency at run time & Energy efficiency & \cite{kim2020}\cite{Mohammed2020}\\
\hline
T10: Use dynamic parameter adaptation & Design parameters that are dynamically adapted based on the input data & Energy efficiency & \cite{Garcia-Martin2021}\\
\hline
T11: Use built-in library functions & Use built-in libraries for ML models if possible & Performance* & \cite{shanbhag2022towards} \\
\hline
\multicolumn{4}{l} {\small{The * means energy efficiency was considered a secondary QA}} \\
\end{tabular}
\label{tactics-algorithm}
\end{table}

\textbf{T6: Choose an energy-efficient algorithm}. Different ML algorithms have different levels of energy consumption and computational power.
For example, the K-nearest neighbor (KNN) algorithm has much lower energy consumption than the ensemble method Random Forest (RF)~\cite{verdecchia2022data}.\footnote{\textbf{Caveat:} in the published version of the paper (\url{https://doi.org/10.1145/3639475.3640111}), this comparison is the other way around, which is wrong. We have corrected the mistake in this postprint.}
High energy consumption does not necessarily mean that the algorithms perform better or achieve higher accuracy levels than low-energy algorithms.
Thus, choosing suitable, energy-efficient algorithms that achieve wanted outcomes can reduce the energy consumption of ML models~\cite{kaack2022aligning}.

\textbf{T7: Choose a lightweight algorithm alternative.} Some algorithms may have lightweight alternatives.
Using these lighter models can have a lower energy consumption without sacrificing other important QAs.
For example, spiking neural networks (SNN) are seen as a lightweight and energy-efficient alternative to convolutional neural networks (CNN).
CNN models can be converted to SNN without a significant loss of accuracy or performance~\cite{sorbaro2020}.

\textbf{T8: Decrease model complexity}. Complex ML models have been shown to have high energy consumption, and therefore scaling down the model complexity can contribute to environmental sustainability.
Simplifying the model structure can lead to faster training and inference times, making it more efficient to deploy and use in real-world applications.
For example, using a simple three-layered Convolutional Neural Network architecture~\cite{morotti2021} and shallower Decision Trees~\cite{abreu2020} has shown to be energy-efficient while still providing high levels of precision.

\textbf{T9: Consider reinforcement learning for energy efficiency}.
Algorithms can be designed to optimize energy efficiency through reinforcement learning.
Reinforcement learning receives feedback on its actions and adjusts its behavior accordingly.
Reinforcement learning models can be used to identify the most energy-efficient options in real time and make informed decisions based on that information~\cite{Mohammed2020, kim2020}.
Additionally, other QAs can also be targeted for optimization, e.g., accuracy or CPU/RAM usage.

\textbf{T10: Use dynamic parameter adaptation}.
Dynamic parameter adaptation means that the hyperparameters of an ML model are dynamically adapted based on the input data, instead of determining the exact parameters values in the algorithm.
For example, \citet{Garcia-Martin2021} used an \textit{nmin} adaptation method for very fast decision trees.
The \textit{nmin} method allows the algorithm to grow faster in those branches where there is more confidence in creating a split, and delaying the split on the less confident branches.
This method resulted in decreased energy consumption.

\textbf{T11: Use built-in library functions}.
Apply built-in library functions in the ML model instead of writing custom implementations.
The existing built-in library functions are usually optimized and well-tested, which is why they may have improved performance and energy efficiency compared to custom-made functions.
For example, these built-in libraries can be used for tensor operations~\cite{shanbhag2022towards}.

\subsection{Model Optimization}
The model optimization category includes all the tactics that are related to the model optimization stage of the ML model development process.
These tactics (T12-T17) are shown in Table \ref{tactics-optimization}.

\begin{table}[ht]
\caption{Green Tactics Related to Model Optimization}
\centering \small 
\begin{tabular}{>{\raggedright\arraybackslash}p{1.9cm}>{\raggedright\arraybackslash}p{3cm}>{\raggedright\arraybackslash}p{1.5cm}>{\raggedright\arraybackslash}p{0.8cm}}
\textbf{Tactic} & \textbf{Description} & \textbf{Target QA} & \textbf{Source} \\
\hline T12: Set energy consumption as a model constraint & Consider energy consumption as one predetermined parameter for optimizing the ML model & Energy efficiency & \cite{Wang2021}\cite{yang2019} \\
\hline T13: Consider graph substitution & Replace energy-intensive model parts with similar, but less energy-consuming parts & Energy efficiency & \cite{Wang2020May} \\
\hline T14: Enhance model sparsity
& Reduce the number of model parameters or set their values to zero & Energy efficiency & \cite{Yang2020a} \\
\hline T15: Consider energy-aware pruning & Prune neural networks starting from the most energy-intensive layer & Energy efficiency & \cite{Yang_2017_CVPR} \\
\hline T16: Consider transfer learning & Use pre-trained ML models for other similar tasks & Energy efficiency & \cite{jayakodi2020}\cite{shanbhag2022towards} \\
\hline T17: Consider knowledge distillation & Use knowledge from a large ML model to train a smaller model & Performance* & \cite{shanbhag2022towards}\cite{yang2019} \\
\hline
\multicolumn{4}{l} {\small{The * means energy efficiency was considered a secondary QA}} \\
\end{tabular}
\label{tactics-optimization}
\end{table}

\textbf{T12: Set energy consumption as a model constraint}.
This tactic sets a predetermined energy consumption threshold for the ML model optimization process.
The optimization takes into account the energy consumption of the model during both the optimization and training phases.
The objective is to train the model in a way that it stays within the specified energy consumption threshold.
This approach views model optimization as an optimization problem, where for instance hyperparameters and the model itself are optimized based on predetermined limits~\cite{Wang2021, yang2019}.

\textbf{T13: Consider graph substitution}.
In the context of deep neural networks (DNN), substitution refers to replacing a large model with a smaller one that performs a similar task.
Energy-aware substitution, however, means replacing energy-intensive nodes of DNNs with less energy-consuming nodes.
For example, \citet{Wang2020May} showed energy savings of 24\% in their study of energy-aware graph substitution.

\textbf{T14: Enhance model sparsity}.
Enhancing the sparsity of an ML model means reducing the number of model parameters or setting their values to zero.
For example, weight sparsification involves identifying and removing unnecessary or less important weights in a neural network.
Enhancing model sparsity decreases the complexity of the model and consequently reduces requirements for storage and memory.
Therefore, it also results in lower power consumption~\cite{yang2020}.

\textbf{T15: Consider energy-aware pruning}.
Pruning is the process of reducing the complexity and size of an ML model by removing unnecessary or less important components, such as weight.
In energy-aware pruning, energy consumption of a neural network is used to guide the pruning process to optimize for the best energy efficiency.
With the estimated energy for each layer in a CNN model, the algorithm performs layer-by-layer pruning, starting from the layers with the highest energy consumption to the layers with the lowest energy consumption.
For pruning each layer, it removes the weights that have the smallest joint impact on the output feature maps~\cite{yang2019}.

\textbf{T16: Consider transfer learning}.
Transfer learning means using knowledge gained from a task (pre-trained model) and transferring it to another similar task.
This is feasible only if there is an existing pre-trained model available for use.
The absence of or reduction in the effort for model training results in savings in energy consumption~\cite{jayakodi2020, shanbhag2022towards}.

\textbf{T17: Consider knowledge distillation}.
Knowledge distillation is a technique where a large, complex model (teacher) is used to train a smaller, simpler model (student).
The goal is to transfer the learned information from the teacher model to the student model, allowing the student model to achieve comparable performance while requiring fewer computational resources~\cite{yang2019, shanbhag2022towards}.

\subsection{Model Training}
Three tactics (T18-T20) are related to ML model training, and are shown in Table \ref{tactics-training}.

\begin{table}[ht]
\caption{Green Tactics Related to Model Training}
\centering \small 
\begin{tabular}{>{\raggedright\arraybackslash}p{1.9cm}>{\raggedright\arraybackslash}p{2.9cm}>{\raggedright\arraybackslash}p{1.6cm}>{\raggedright\arraybackslash}p{0.8cm}}
\textbf{Tactic} & \textbf{Description} & \textbf{Target QA} & \textbf{Source} \\
\hline T18: Use quantization-aware training & Convert high-precision data types to lower precision during training & Accuracy* &~\cite{sorbaro2020, Kim2021} \\
\hline T19: Use checkpoints during training & Use checkpoints to avoid a knowledge loss in case of a premature termination & Recoverability* &~\cite{shanbhag2022towards} \\
\hline T20: Design for memory constraints & Consider possible memory constraints during training & Recoverability* &~\cite{shanbhag2022towards} \\
\hline
\multicolumn{4}{l} {\small{The * means energy efficiency was considered as a secondary QA}} \\
\end{tabular}
\label{tactics-training}
\end{table}

\textbf{T18: Use quantization-aware training}.
Quantization-aware training is a technique used to train neural networks to convert data types to lower-precision ones.
The idea is to use fixed-point or integer representations instead of the more commonly used higher precision floating-point representations.
This improves the performance and energy efficiency of the model in federated learning~\cite{sorbaro2020, Kim2021}.

\textbf{T19: Use checkpoints during training}.
Training is an energy-intensive stage of the ML model life cycle, which may take long periods of time.
Sometimes, a failure or a hardware error can terminate the training process before it is completed.
In those cases, the training process has to be started from the beginning and all progress is lost.
The use of checkpoints, however, can save progress in regular intervals and in case of a premature termination, the training process can continue from the last checkpoint~\cite{shanbhag2022towards}.

\textbf{T20: Design for memory constraints}.
Model training requires memory, and sometimes memory leaks and OOM (out of memory) errors may occur during that process.
If that happens, the knowledge gained during the prior training process is lost.
By considering memory availability constraints and addressing possible OOM exceptions, the model training pipeline can be designed to operate within the available memory limits.
It reduces the likelihood of errors and prevents unnecessary energy consumption~\cite{shanbhag2022towards}.

\subsection{Deployment}
The deployment category contains tactics related to deploying an ML-enabled system such that it is more environmentally sustainable.
These tactics (T21-T27) are shown in Table \ref{tactics-depl}.

\begin{table}[ht]
\caption{Green Tactics Related to Model Deployment}
\centering \small 
\begin{tabular}{>{\raggedright\arraybackslash}p{1.9cm}>{\raggedright\arraybackslash}p{2.9cm}>{\raggedright\arraybackslash}p{1.6cm}>{\raggedright\arraybackslash}p{0.8cm}}
\textbf{Tactic} & \textbf{Description} & \textbf{Target QA} & \textbf{Source} \\
\hline T21: Consider federated learning & Train the model and store data in decentralized devices & Energy efficiency &~\cite{Kim2021}\\
\hline T22 Use computation partitioning & Divide computations between a client and a cloud server & Energy efficiency&~\cite{manasi2020}\\
\hline T23: Apply cloud fog network architecture & Use an architecture in which the models are processed between
end devices and cloud & Energy efficiency &~\cite{yosuf2021}\\
\hline T24: Use energy-efficient hardware & Use energy-efficient, ML-suitable hardware & Energy efficiency &~\cite{kaack2022aligning}\\
\hline T25: Use power capping & Set energy consumption limits for hardware& Energy efficiency &~\cite{krzywaniak2022gpu}\\
\hline T26: Use energy-aware scheduling & Dynamically optimize the scheduling of ML tasks & Resource utilization* &~\cite{sun2020}\\
\hline T27: Minimize referencing to data & Avoid unnecessary read and write data operations & Energy efficiency &~\cite{shanbhag2022towards}\\
\hline
\multicolumn{4}{l} {\small{The * means energy efficiency was considered a secondary QA}} \\
\end{tabular}
\label{tactics-depl}
\end{table}

\textbf{T21: Consider federated learning}.
Federated learning is an ML approach that aims to train a shared ML model on decentralized devices.
Instead of sending raw data to a central server, federated learning trains the model directly on the devices where the data is generated, such as mobile phones or edge devices.
Only the trained model or updated model parameters are sent to a central server~\cite{Kim2021}.
Federated learning decreases the resources needed for transferring large amounts of data to a central server, which results in improved energy efficiency.

\textbf{T22: Use computation partitioning}.
Computation partitioning is the process of dividing the computations of a CNN between a mobile client and a cloud server.
The goal is to optimize energy consumption and efficiency.
The NeuPart framework~\cite{manasi2020} is an example of a partitioning approach.
NeuPart divides computational tasks between the mobile device (client) and the remote server or data center (cloud) in real time based on energy consumption.
By offloading computationally intensive tasks to the cloud and executing lighter tasks locally, NeuPart resulted in significant energy savings of up to 52\% in cloud-based computations~\cite{manasi2020}.

\textbf{T23: Apply cloud fog network architecture}.
Edge devices are usually connected to distant cloud services.
However, bringing the cloud closer to edge devices could be more energy efficient.
A cloud fog network (CFN) is one way to achieve that~\cite{yosuf2021}.
CFN supports an architecture where deep neural network models are processed in servers between end-devices and the cloud.
For example, \citet{yosuf2021} present a CFN architecture that consists of four layers: IoT end devices, Access Fog (AF), Metro Fog (MF) and Cloud Datacenter (CDC).
This architecture led to a 68\% reduction in power consumption when compared to a traditional cloud data center architecture on average.

\textbf{T24: Use energy-efficient hardware}.
The emissions of ML are related to used hardware.
This is why using energy-efficient hardware to run ML models can reduce their power consumption.
Energy-efficient hardware can include low-energy components.
For example, the Tensor Processing Units (TPUs) developed by Google are seen as an energy-efficient alternative to CPUs and GPUs~\cite{kaack2022aligning}.

\textbf{T25: Use power capping}.
Power capping is a technique used to limit the amount of power consumed by a device or system, such as a CPU, GPU, or server.
It involves setting a maximum power consumption threshold for a device, and dynamically adjusting the power usage to ensure that it stays below that threshold.
This is typically done to manage the power consumption and heat dissipation of a device, and to prevent it from exceeding the power budget of a data center or other power-limited environment.
Restricting the use of GPU resources can lead to reduced performance and longer execution times, but in certain configurations, it can also result in a significant reduction in energy consumption (up to 33\%) with a moderate impact on performance~\cite{krzywaniak2022gpu}.

\textbf{T26: Use energy-aware scheduling}.
Energy-aware scheduling refers to a strategy that optimizes the scheduling of ML tasks.
It dynamically schedules tasks or processes based on the current energy requirements and system conditions.
The objective of an energy-aware dynamic scheduling policy is to make efficient use of available computational resources while still meeting energy budgets~\cite{sun2020}.

\textbf{T27: Minimize referencing to data}.
ML models require reading and writing enormous amounts of data in the ML workflow.
Reading data means retrieving information from storage, while writing data means storing or updating the information.
These operations may increase unnecessary data movements and memory usage, which influence the energy consumption of computing.
To avoid non-essential referencing of data, data read and write operations must be designed carefully~\cite{shanbhag2022towards}.

\subsection{Management}
The management category includes tactics for managing the ML-enabled system and ML model after their deployment.
Only three tactics (T28-T30) fall under this category and are shown in Table \ref{tactics-management}.

\begin{table}[ht]
\caption{Green Tactics Related to Management}
\centering \small 
\begin{tabular}{>{\raggedright\arraybackslash}p{1.5cm}>{\raggedright\arraybackslash}p{2.9cm}>{\raggedright\arraybackslash}p{1.7cm}>{\raggedright\arraybackslash}p{0.8cm}}
\textbf{Tactic} & \textbf{Description} & \textbf{Target QA} & \textbf{Source} \\
\hline T28: Use informed adaptation & Adapt the model based on informed concept shift & Energy efficiency & \cite{poenaruretrain} \\
\hline T29: Retrain the model if needed & In case of concept shift, retrain the existing ML model instead of building a new one & Accuracy* & \cite{poenaruretrain}
 \\
\hline T30: Monitor computing power & Monitor computing power of an ML model in the long-term & Energy efficiency & \cite{Cao}\cite{Kumar2020}\\
\hline
\multicolumn{4}{l} {\small{The * means energy efficiency was considered a secondary QA}} \\
\end{tabular}
\label{tactics-management}
\end{table}

\textbf{T28: Use informed adaptation}.
ML models may experience drift that affects their functionality.
In these cases, the models must be adapted to deal with the drift.
Informed adaptation refers to a method of adapting the ML model only when drift is detected.
Therefore, the frequency of adaptation is smaller than in blind, periodic adaptation.
Informed adaptation reduces unnecessary adaptations, which consequently saves energy~\cite{poenaruretrain}.

\textbf{T29: Retrain the model if needed}.
Retraining a model refers to the process of updating or modifying an existing ML model.
In the long term, concept drift may affect the accuracy of existing ML models.
Retraining the model, by for example training it again with new data, is better than building it again from scratch in terms of sustainability~\cite{poenaruretrain}.

\textbf{T30: Monitor computing power}.
Estimating and calculating the energy footprint of an ML model can help to reduce its computational power consumption.
Monitoring the energy consumption of an ML model over the long term helps to identify those components where energy is being inefficiently utilized.
This can serve as a starting point for making improvements to reduce energy consumption.
There has been a lack of easy-to-use tools to do that, but recently researchers have provided frameworks on how to estimate or calculate the energy footprint of ML-enabled systems~\cite{Cao, Kumar2020}.

\subsection{Targeted Quality Attributes}
The majority of the tactics (21 out of 30) unsurprisingly aim to improve energy efficiency as their primary QA.
Energy efficiency was mostly measured by the energy consumption savings achieved with the tactic.
However, there were also some tactics that aimed primarily at other QAs, with energy efficiency being improved as a side effect.
The description of these QAs and the number of tactics targeting them can be found in Table~\ref{qas}.
One example was performance (3 out of 30), which was usually related to the time or throughput related to model training or inference.
If the runtime or computational intensity can be reduced, this sometimes also influences energy efficiency positively.
Additionally, some tactics had accuracy as their primary QA (3).
While energy efficiency and accuracy are often regarded as a trade-off~\cite{yang2019}, our collection contains several tactics that try to increase both simultaneously.
Lastly, recoverability (2) and resource utilization (1) appeared for a small number of tactics.

Overall, 17 tactics (T1-T6, T8, T10, T12-T15, T17, T22, T23, T25 and T26) were evaluated in experimental settings to provide evidence for their impact on these QAs.
Most of the papers also provided evaluations of possible trade-offs with other QAs.
These trade-offs include, for example, accuracy and latency.
The other remaining tactics (13) were more along the lines of experience-based suggestions to improve QAs and environmental sustainability, without rigorous evaluations.
While the provided argumentation was convincing, future work needs to provide empirical evidence to quantify the impact.

\begin{table}[ht]
\caption{Target Quality Attributes of the 30 Green Tactics}
\centering \small 
\begin{tabular}{>{\raggedright\arraybackslash}p{2cm}>{\raggedright\arraybackslash}p{4.3cm}>{\raggedright\arraybackslash}p{1.3cm}}
\hline
\textbf{QA} & \textbf{Description} & \textbf{\#} \\
\hline Energy efficiency & The ability to accomplish a task while minimizing energy consumption & 21\\
\hline Performance & The efficiency with which a task is achieved (e.g., speed, stability) & 3\\
\hline Accuracy & The level of how accurately the algorithm performs specified tasks.
& 3\\
\hline Recoverability & The ability to restore and resume normal operations after a failure & 2\\
\hline Resource utilization & The ability to use and allocate resources efficiently & 1\\
\hline
\end{tabular}
\label{qas}
\end{table}

\subsection{Scope of Architectural Tactics}
The majority of tactics in our collection (20 of 30) are associated with low-level phases of the ML development life cycle, namely data collection and processing, algorithm design, model optimization, and model training.
In essence, these tactics are targeted at improving model quality instead of system quality.
For example, tactics like \textit{Reduce the number of data features} (T3) or \textit{Decrease model complexity} (T8) could also be applied without a complete software system, i.e., for the training of a single ML model that is never integrated into a larger system.
However, when these tactics are applied systematically at scale and continuously, they have strong architectural implications and significance.
For example, although accuracy could be considered a model quality concern, the system actions taken in response to reduced accuracy as a trade-off for energy efficiency are architectural in nature.
Moreover, some tactics from the early life cycle phases have a profound influence on architectural elements in the system, e.g., \textit{Consider reinforcement learning for energy efficiency} (T9).
For the categories \textit{deployment} and \textit{management}, the architectural significance of the tactics is more obvious, e.g., for \textit{Consider federated learning} (T21), \textit{Apply cloud fog network architecture} (T23), or \textit{Use informed adaptation} (T28).
All in all, the collection of tactics provides practitioners with holistic, architecture-centric means to improve the environmental sustainability of ML-enabled systems in all life cycle phases.

\section{Discussion}
\label{s:discussion}
The collection of green architectural tactics obtained from our study provides guidance for architecting environmentally sustainable ML-enabled systems, and, as such, realize the societal promise that is expected of ML: minimizing its energy footprint.
In the following, we report the main observations and reflections about our study.

\textbf{Due to their diversity, tactic selection requires domain-specific expertise.}
As mentioned earlier, the tactics in Fig.~\ref{data-tactics} cannot be generalized to all use cases.
For instance, some tactics are suitable for a specific algorithm type (e.g., \textit{Consider graph substitution} (T13) is only applicable to neural networks), while others are conditional to specific requirements (e.g., \textit{Consider transfer learning} (T16) requires a pre-trained model).
Therefore, our catalog of tactics should not be interpreted as strict rules, but rather as recommendations or available techniques for the architecture design of energy-efficient ML-enabled systems.
In perspective, we believe that it would be beneficial for ML practitioners to have a catalog of tactics for specific use cases, such as using deep learning and its software architecture implications.

\textbf{Most tactics are model-related rather than focusing on the full architecture of ML-enabled systems.}
An important observation also discussed in the focus group is that most tactics found in our study focus on the ML model rather than the architecture of ML-enabled systems, i.e., a specific component rather than substantial parts of the architecture.
We argue that this is because the field is still maturing, and reusable architecture knowledge about ML-enabled systems is still in the making.
Furthermore, a related open problem is being able to separate the energy efficiency of model development from the energy efficiency of the ML-enabled systems that use such models, which may not even be fully achievable.

\textbf{Studying the interaction between energy efficiency and other QAs may accelerate the creation of architecture knowledge for ML-enabled systems.}
Even though our study focuses on energy efficiency as its primary target QA, we observed that many of the reviewed papers also evaluated the tactic's impact on accuracy.
Historically, accuracy and energy efficiency have been seen as a trade-off: high accuracy requires intensive training, which in turn is energy-demanding.
Naturally, if accuracy decreases significantly due to an increased energy-efficiency target, we run the risk that the models are not usable anymore because they do not produce the desired outcomes. 
Therefore, accuracy is an important QA for ML models and ML-enabled systems, even though it is not always the primary target of a tactic.
However, recent research also highlights that, in many cases, energy consumption of ML models can be reduced \textit{without} substantial reduction in accuracy~\cite{verdecchia2022data,XuYinlena2023,Yarally2023,delRey2023}.
In general, given that energy consumption has become a major concern only recently and that ML-enabled systems are extremely energy demanding, we argue that future research should investigate possible trade-offs between energy efficiency and other QAs.
Analyzing interactions between multiple QAs could provide important insights into the design of ML-enabled systems.

\section{Conclusions and Next Steps}
\label{s:conclusion}
This paper provides a catalog of 30 green architectural tactics for ML-enabled systems organized in 6 categories, namely \textit{data-centric}, \textit{algorithm design}, \textit{model optimization}, \textit{model training}, \textit{deployment}, and \textit{management}.
The tactics represent available techniques for designing energy-efficient ML-enabled systems.
We integrated the collection into the Archive of Awesome and Dark Tactics.\footnote{\url{https://s2group.cs.vu.nl/AwesomeAndDarkTactics/catalog}}
For transparency and reusability, we also provide a Zenodo repository.\footnote{\url{https://doi.org/10.5281/zenodo.8349208}}

Despite the growing understanding of the environmental impacts of ML, there is still no consensus on how to best achieve sustainability.
Our study serves as a starting point for further research about green architectural tactics for ML-enabled systems.
Future research is necessary to validate and extend the results of this study, and to explore more generalized tactics applicable to different ML algorithms and ML-enabled system concerns.
Furthermore, more research is required to evaluate the effectiveness of several green architectural tactics in practice.
In line with this, we plan to re-engineer existing open-source ML-enabled systems based on the tactics and to measure energy efficiency to compare the original versus the modernized system.
As an alternative, we may conduct a case study where practitioners develop an ML-enabled system using the tactics catalog.
Based on this, the usage of concrete tactics is analyzed, experiences are documented, and the catalog is refined.
Finally, with the energy crisis and the explosion of ML applications in all sectors, energy efficiency is gaining traction as an important QA.
Accordingly, it is important to create tactics dedicated to raising awareness of the energy footprint of such systems.

\begin{acks}
We kindly thank our three focus group experts for their valuable time and feedback!
Additionally, we thank Iffat Fatima (VU Amsterdam) for her support with integrating the tactics into the AADT.
Lewis and Ozkaya's work was supported by the Department of Defense under Contract No. FA8702-15-D-0002 with Carnegie Mellon University for the operation of the Software Engineering Institute, a federally funded research and development center (DM23-2363).
Muccini's work was supported by the PNRR ICSC National Research Centre for High Performance Computing, Big Data and Quantum Computing (CN00000013), under the NRRP MUR program funded by the NextGenerationEU.
This research was supported by ExtremeXP, a project co-funded by the European Union Horizon Programme under Grant Agreement No. 101093164.
\end{acks}

\bibliographystyle{ACM-Reference-Format}
\bibliography{references}

\end{document}